\documentclass[epj,nopacs]{svjour}
 \usepackage{mathrsfs,cite}
\usepackage{amsmath, txfonts}
\usepackage{graphicx,bm,booktabs,bm}
\usepackage{longtable,lscape}
\usepackage{overpic,multirow,color}
\usepackage{ulem}
\usepackage[colorlinks,
            citecolor=blue,
            anchorcolor=green,
            menucolor=orange,
            linkcolor=red,
            filecolor=red,
            runcolor=pink,
            urlcolor=blue,
            frenchlinks=red]{hyperref}

\allowdisplaybreaks[4]

\graphicspath{{Figures/}} %

\def\zc{Z_c}
\def\zcp{Z_c^\prime}

\begin{document}
\title{Production of $Z_c(3900$) and $Z_c(4020)$ in $B_c$ decay}
\author{Qi Wu\inst{1}
\and Dian-Yong Chen\inst{1}
\thanks{\emph{chendy@seu.edu.cn (Corresponding author)}}%
\and Xue-Jia Fan\inst{2}
\and Gang Li\inst{2}
\thanks{\emph{gli@qfnu.edu.cn (Corresponding author)}}%
}
%
%
\institute{{School of Physics, Southeast University, Nanjing 210094,  China}\and
{College of Physics and Engineering, Qufu Normal University, Qufu 273165, China}}

\date{Received: date / Revised version: date}
%
\abstract{
In the present work, we propose to search the charmonium-like states $Z_c(3900$) and $Z_c(4020)$ in the $B_c$ decay. In an effective Lagrangian approach, the branching ratios of $B_c^+ \to Z_c(3900)^+ \pi^0$ and $B_c^+\to Z_c (4020)^+ \pi^0$ are estimated to be of order of $10^{-4}$ and $10^{-7}$, respectively. The large production rate of $Z_c(3900)$ could provide an important source of the production of $Z_c(3900)$ from the semi-exclusive decay of $b$-flavored hadrons reported by D0 Collaboration, which can be tested by the exclusive measurements in LHCb.
} 
\maketitle

\section{Introduction}
\label{sec:introduction}
The charmonium-like state $Z^\pm_c(3900)$ (abbreviate to $Z_c$ here and after) was first observed in the $J/\psi\pi^\pm$ invariant mass spectrum of $e^+ e^-\rightarrow\pi^+ \pi^- J/\psi$ by BESIII and Belle Collaborations \cite{Ablikim:2013mio, Liu:2013dau}, and then confirmed in $D^\ast \bar{D}$ invariant mass spectrum of $e^+ e^- \to \pi^\pm(D\bar{D}^\ast)^\mp$ process \cite{Ablikim:2013xfr}. The spin and parity quantum numbers of the $Z_c$ had been determined to be $J^P=1^+$ by the partial wave analysis of the process $e^+ e^-\rightarrow\pi^+\pi^- J/\psi$ \cite{Collaboration:2017njt}. As a partner of $Z_c$, $Z_c(4020)$ (abbreviate to $Z_c^\prime$ here and after), was discovered in the $h_c\pi^\pm$ invariant mass spectrum of $e^+ e^-\rightarrow\pi^+ \pi^- h_c$ \cite{Ablikim:2013wzq} and confirmed in the $D^\ast \bar{D}^\ast$ invariant mass spectrum of $e^+ e^- \to (D^\ast \bar{D}^\ast)^\pm \pi^\mp$ process \cite{Ablikim:2013emm} by the BESIII Collaboration.

The decays $Z^\pm_c(3900)\rightarrow J/\psi\pi^\pm$ and $Z^\pm_c(4020)\rightarrow h_c\pi^\pm$ indicates $Z^+_c$ and $Z^{\prime+}_c$ most likely contains at least four constitute quarks. Thus these two states can be good candidates of tetraquark states.  In Ref. \cite{Qiao:2013dda}, $\zc$ was considered as the charm counterpart of $Z_b(10610)$ with $[cq][c\bar{q}]$ tetraquark configuration, and the mass estimated by using QCD sum rule (QSR) was consistent with the observed one of $\zc$ within the errors. The authors in Refs.\cite{Wang:2013vex,  Wang:2013llv, Wang:2013exa, Agaev:2017tzv, Zhao:2014qva} assigned both $\zc$ and $\zcp$ as a diquark-antidiquark tetraquark state with $J^{P}=1^+$, but the estimation in Ref. \cite{Zhao:2014qva} indicated that the lowest axial-vector tetraquark sates dominantly decay into $J/\psi \pi$, while their open charm decay modes were strongly suppressed , which is contrast with the experimental measurements.  By reanlaysing the experimental data of $J/\psi \pi$ invariant mass spectrum from the BESIII and Belle collaboration, the authors indicated that the data may contain another resonance and this resonance as well as $Z_c(3900)$ could be tetraquark states  \cite{Faccini:2013lda}.  There are also some estimation based on constitute quark model, such as, QCD confining model based on SU(3) flavor symmetry \cite{Zhu:2016arf}, color flux-tube model with a four-body confinement potential \cite{Deng:2014gqa}, relativized diquark model \cite{Anwar:2018sol}, and  in Ref. \cite{Patel:2014vua}, the structures observed in $J/\psi \pi$ and $D^\ast \bar{D}$ were not the same, the former one could be interpreted as a $[Q\bar{q}][\bar{Q}q]$ tetraquark state, while the latter one was considered as a  $[Qq][\bar{Q} \bar{q}]$ tetraquark state.

It should be notice that the observed masses of $Z_c$ and $Z_c^\prime$ are in the vicinity of $D^\ast \bar{D}$ and $D^\ast \bar{D}^\ast$ thresholds, respectively, thus, they have been considered as the molecular or resonance states resulted from the $D^\ast \bar{D}^{(\ast)}$ interactions. The estimations in a one-boson-exchange model indicated that $\zc$ and $\zcp$ could be molecular states composed of $D\bar{D}^\ast +c.c$ and $D^\ast \bar{D}^\ast$, respectively \cite{Liu:2008tn, Liu:2008fh,Sun:2012zzd, He:2013nwa}, where the long range pion exchange plays an important role. By using a $D^\ast \bar{D}^\ast $ interpolating current within QSR, the mass of the molecular state was consistent with $\zcp$ \cite{Chen:2013omd}.  The decay behaviors of $\zc$ and $\zcp$ were estimated in a phenomenological Lagrangian approach \cite{Gutsche:2014zda, Chen:2015igx}. In Refs. \cite{Li:2013xia,Li:2014pfa}, the hidden charm decays of  $\zc$ and $\zcp$ were estimated based on heavy quark symmetry. The estimations in Ref.  \cite{Xiao:2018kfx} indicated that the hidden charm decays of $\zc$ and $\zcp$ could be understood by the final states interactions, which indicated that these two charmonium-like states could be molecular candidates. In Ref. \cite{Wu:2016ypc}, the charmless decays of  $\zc$ and $\zcp$ were investigated in an effective Lagrangian approach. With final states interaction effect, a pole corresponding to $\zc$ in the complex energy plane in different Riemann sheets could be found, which supported $\zc$ as a $D^\ast \bar{D}+c.c$ molecular state \cite{Gong:2016hlt}.  By considering the couple channel effect of $\pi J/\psi-\bar{D}^*D$ interaction, $\zc$ was found to be a virtual state \cite{He:2017lhy}.

Besides the tetraquark and molecular scenarios, there are also some nonresonance interpretations, such as cusp effect \cite{Liu:2013vfa,Swanson:2014tra} and initial single pion emission mechanism \cite{Chen:2011xk, Chen:2013coa}, in which $Z_c$ and $Z_c^\prime$ were considered as the kinetic effects due to $D^\ast \bar{D} +h.c$ and $D^\ast \bar{D}^\ast $  interactions, respectively. In addition, the Lattice QCD estimations also indicated that $\zc$ is not a conventional resonance but a threshold cusp \cite{Ikeda:2017mee, Ikeda:2016zwx}.

In addition to the mass spectrum and decay behaviors, the production process can also provide some important information of the internal structures of $Z_c^{(\prime)}$, thus investigating the productions of $Z_c^{(\prime)}$ in different processes are very critical.  The observed process $Y(4260) \to \zc \pi$ were investigated in Ref. \cite{Chen:2016byt}, where $Y(4260)$ and $\zc$ were treated as $D\bar{D}_1+c.c$ and $D^\ast \bar{D}+c.c$ molecular states, respectively. In Ref. \cite{Lin:2013mka}, a large cross sections for $\gamma p \to \zc^+ n$ was predicted , however, the COMPASS Collaboration found no evidence of $\zc$, and an upper limit for the ratio $\mathcal{B}(\zc \to J/\psi \pi)  \times \sigma(\gamma N\to \zc N)/\sigma (\gamma N \to J/\psi N)$ were measured to be $3.7 \times 10^{-3}$ at $90\%$ credit level. The cross section for the Pion-induced production of $Z_c$ off a nuclear target was estimated in Ref. \cite{Huang:2015xud},  where the nuclear target will enhance the meson productions. Recently, $\zc$ is observed in the $J/\psi \pi$ invariant mass spectrum of the semi-inclusive weak decays of b-flavored hadron \cite{Abazov:2018cyu}. However, in the exclusive $B$ decay process, such as $B \to K J/\psi \pi \pi$ \cite{Choi:2003ue} and $B\to K D^\ast \bar{D}$ \cite{Adachi:2008sua} , no signal of $\zc$ is observed. Then, searching for the source of  $\zc$ production in the semi-inclusive b-flavor hadron decay will be interesting. In addition, the production of $\zc$ and $\zcp$ are similar in electron-positron annihilation process, weather such kind of similarity still holds in $b-$flavored hadron decay is a question needed to be answer. Comparing to $B_s^0$, $B_c^+$ should more easily be detected experimentally, thus, in the present work, we investigate $\zc^{(\prime)}$ production in the $B_c^+$ decays.

\begin{figure}[t]
\setlength{\tabcolsep}{2mm}{
\begin{tabular}{ccc}
  \centering
 \includegraphics[width=4cm]{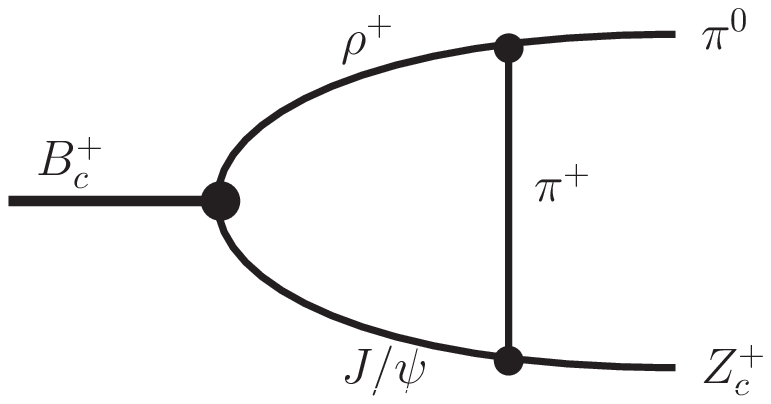}&
 \includegraphics[width=4cm]{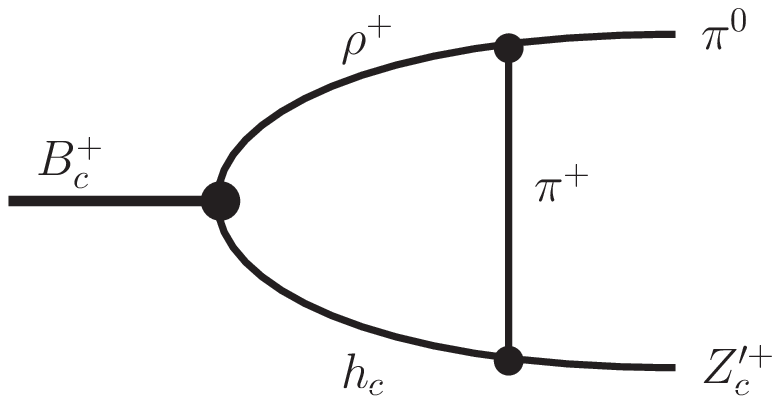}\\
 $(a)$ & $(b)$ \\
 \end{tabular}}
  \caption{ Sketch diagrams for $B_c\rightarrow Z_c\pi$ (diagram \textbf{a}) and $B_c\rightarrow Z^\prime_c\pi$  (diagram \textbf{b}).}\label{fig:feyn-Zb-1}
\end{figure}

The experimental measurements indicate that $Z_c^{(\prime)}$ dominantly decay into a pair of charmed meson or a charmonium and a pion. Thus, we have two choice to connect $Z_c^{(\prime)}$ with the mother particle $B_c^+$, one is through the charmed meson loop and the other is via the loop formed by a charmonia and a light meson.  Theoretical estimations of the branching ratios of two charm meson decays of $B_c^+$ are of order of $10^{-6}$ \cite{Rui:2012qq,Kiselev:2002vz,Ivanov:2002un,Ivanov:2006ni}, and the LHCb collaboration have measured the decays of $B_c^+ \rightarrow D^{(\ast)+}_{(s)}\bar{D}^{(\ast)0}$ and $B_c^+ \rightarrow D^{(\ast)+}_{(s)}D^{(\ast)0}$ and no signals was found \cite{Aaij:2017gon}.  Thus, the scenario of $Z_c^{(\prime)}$ productions from $B_c^+$ decay via a charmed meson loop should be rule out.  As for the second choice, one should notice that  $B^{+}_c\rightarrow c\bar{c}+\rho^{+}$ is a CKM enhanced mode with $(\Delta b=1, \Delta C=1, \Delta S=0)$ and $\mathcal{B}(B^+_c\rightarrow J/\psi/h_c+\rho^+)$ were predicted to be of order of $10^{-3}$ \cite{Ivanov:2006ni,Chang:2001pm}. The $ \rho$ meson dominantly decays into a pair of pion, thus, there should exists meson loops composed of $J/\psi\rho/h_c\rho$ in the decay channel $B_c\rightarrow Z_c\pi/Z^\prime_c\pi$ as presented in Fig. \ref{fig:feyn-Zb-1}. Furthermore, all the involved mesons in the loops could be on-shell and the contributions from such loop will be further enhanced, thus $B_c \to Z_c^{(\prime)} \pi$ might be a gold channel of $Z_c^{(\prime)}$ production from $b$ flavor mesons.

This work is organized as follows. After introduction, we present formula of decays of $B_c \to c\bar{c} \rho $ and the amplitudes of the meson loop contributions to $B_c\rightarrow c\bar{c}\rho\rightarrow Z^{(\prime)}_c\pi$  in Sec. ~\ref{sec:Sec2}. The numerical results and discussions are given in Sec.~\ref{sec:results}, and Sec.~\ref{sec:summary} is devoted to a short summary.

\section{Theoretical framework and the decay amplitudes}
\label{sec:Sec2}

The weak decays of $B_c \to (c\bar{c}) \rho $ have been well investigated in the literatures \cite{Ivanov:2006ni,Chang:2001pm}.  For the completeness of this paper, we just present a short review of the decay amplitudes of $B_c \to (c\bar{c}) \rho $, which will be used in our following estimations. At the quark level, the process happened in $B^{+}_c\rightarrow c\bar{c}+\rho^{+}$ is the weak decay $\bar{b}\rightarrow \bar{c}u\bar{d}$ and the charm quark as a spectator. The effective Hamiltonian related to $B^{+}_c\rightarrow c\bar{c}+\rho^{+}$ is \cite{Ivanov:2006ni}
\begin{eqnarray}
\mathcal{H}_{eff}&=&\frac{G_F}{\sqrt{2}}V_{cb}V^{\dag}_{ud}[c_1(\bar{c}b)_{V-A}(\bar{d}u)_{V-A}+c_2(\bar{d}b)_{V-A}(\bar{c}u)_{V-A}]
\nonumber\\
&&+\mathrm{h.c.},
\label{Eq:Heff}
\end{eqnarray}
where the subscript $V-A$ denotes the left-chiral current $\gamma^\mu(1-\gamma^5)$, $c_1$ and $c_2$ is the Wilson coefficients.

The decay amplitudes for the non-leptonic decays can be formulated into the three factors: the non-leptonic decay constants, the weak current matrix elements and the relevant coefficients in the combinations.  In particular, the non-leptonic decay constant are defined by hadronic matrix elements
\begin{eqnarray}
\langle0|\mathcal{J}_\mu|\rho(k,\varepsilon)\rangle=f_{\rho}\varepsilon_\mu m_\rho,
\end{eqnarray}
where $f_\rho$ stands for leptonic decay constant of $\rho$ meson, $\varepsilon_\mu$ denotes the polarization of the vector meson, and the current $\mathcal{J}_\mu$ is $\mathcal{J}_\mu=\bar{q}_1\gamma_\mu(1-\gamma_5)\bar{q}_2$. The transition matrix element between $B_c$ and (pseudo-)vector meson can be expressed in the combination of four form factors, $V$, $A_{0,\pm}$, which are the function of the square of the transfer momentum between $B_c$ and vector meson. The transition matrix element could be expressed as \cite{Wang:2008xt,Issadykov:2018myx, Wang:2009mi,Rui:2017pre,Rui:2014tpa}

\begin{eqnarray}
&&\langle (c\bar{c})_{J=1}(p_2,\epsilon)  \left| \mathcal{J}^{\mu}\right | B_c^+(p) \rangle=\epsilon_\nu^\ast  \big[ i\varepsilon^{\mu\nu\alpha\beta}P_\alpha Q_\beta V(Q^2)
\nonumber\\
 &&\hspace{1cm}-g^{\mu\nu}(P\cdot Q)A_0(Q^2)+P^\mu P^\nu A_+(Q^2)
+Q^\mu P^\nu A_-(Q^2) \big], \label{Eq:TM}\nonumber \\
\end{eqnarray}
where $(c\bar{c})_{J=1}$ indicates $J/\psi$  and $h_c$ here and after and  $P_\mu=(p+p_2)_\mu$ and $Q_\mu=(p-p_2)_\mu$ \footnote{Actually, the expressions of the transition matrix in Refs. \cite{Wang:2008xt,Issadykov:2018myx, Wang:2009mi} is not exactly the same as Eq.  (\ref{Eq:TM}). As for $B_c \to J/\psi$, the transition matrix defined in Refs. \cite{Wang:2008xt,Issadykov:2018myx} is,
\begin{eqnarray}
&&\langle J/\psi(p_2,\epsilon^\ast)|\mathcal{J}^\mu|B_c(p)\rangle = \frac{\epsilon^\ast_\nu}{m_{B_c}+m_{J/\psi}}\big[i\varepsilon^{\mu\nu\alpha\beta}P_\alpha Q_\beta V(Q^2)
\nonumber\\
&&\hspace{1cm}-g^{\mu\nu}(P\cdot Q)A_0(Q^2)+P^\mu P^\nu A_+(Q^2)
+Q^\mu P^\nu A_-(Q^2)\big],
\label{Eq:tran1}
\end{eqnarray}
and the one of $B_c\to h_c$ presented in Ref. \cite{Wang:2009mi} is,
\begin{eqnarray}
&&\langle h_c(p_2,\epsilon^\ast)|\mathcal{J}^\mu|B_c(p)\rangle = \frac{1}{m_{B_c}-m_{h_c}}\varepsilon^{\mu\nu\alpha\beta}\epsilon^\ast_\nu P_\alpha Q_\beta V(Q^2)
\nonumber\\
&&\hspace{1cm}-i(m_{B_c}-m_{h_c})(\epsilon^{\ast\mu}-\frac{\epsilon^\ast\cdot P}{Q^2}Q^\mu)A_+(Q^2) -i \big[-\frac{\epsilon^\ast\cdot P}{m_{B_c}-m_{h_c}}P^\mu
\nonumber\\
&&\hspace{1cm}+(m_{B_c}+m_{h_c}) \frac{\epsilon^\ast\cdot P}{Q^2}Q^\mu \big]A_-(Q^2)
-i2m_{h_c}\frac{\epsilon^\ast\cdot P}{Q^2}Q^\mu A_0(Q^2).
\label{Eq:tran2}
\end{eqnarray}
One can find the expressions of the transition matrix in Eqs. (\ref{Eq:tran1}) and (\ref{Eq:tran2}) can be reduced to the form of Eq. (\ref{Eq:TM}) by absorbing some constant factors into the form factors.
}. The concrete expression of the form factors $V,\ A_{0,\pm}$ will be discussed in the following section.

With Eqs. (\ref{Eq:Heff})-(\ref{Eq:TM}), one can get the decay amplitude of $B_c \to \rho (c\bar{c})_{J=1}$,

\begin{eqnarray}
\mathcal{M}[B^{+}_c(p)\rightarrow\rho^{+}(p_1)(c\bar{c})_{J=1} (p_2)]
\nonumber\\
\equiv \mathcal{A}^{\mu \nu}(P,Q) \epsilon_\mu(p_1) \epsilon_\nu(p_2) , \label{Eq:AmpV1}
\end{eqnarray}
where $\mathcal{A}^{\mu \nu}(P,Q)$ is defined as
\begin{eqnarray}
\mathcal{A}^{\mu \nu}(P,Q)
&=&\frac{G_F}{\sqrt{2}}V_{cb}V^{\dag}_{ud}a_1 f_{\rho}m_\rho \big[i\varepsilon^{\mu\nu\alpha\beta}P_\alpha Q_\beta V(Q^2)
\nonumber\\
&&-g^{\mu\nu}P\cdot Q A_0(Q^2)+P^\mu P^\nu A_+(Q^2)
\nonumber\\
&&+Q^\mu P^\nu A_-(Q^2) \big],\label{Eq:AmpV1}
\end{eqnarray}
where the Fermi coupling constant $G_F=0.16637 \times 10^{-5}$ GeV$^{-2}$, the CKM matrix elements $|V_{cb}|=0.0405$, $|V_{ud}|=0.974$ \cite{Patrignani:2016xqp}, and $a_1=c_1+\frac{1}{N_c}c_2$ with $N_c=3$ to be the number of color. In the present work, we take  $a_1=1.14$ \cite{Ivanov:2006ni}. The decay constant of $\rho$ meson is taken as $f_\rho=221$ MeV \cite{Cheng:2004ru,Lu:2005mx}.

As shown in Fig.\ref{fig:feyn-Zb-1},  the initial $B_c$ and final $\zc^{(\prime)} \pi$ could be connected by meson loops composed by  $(c\bar{c})_{J=1} \rho $. In the present work, we estimate the meson loop contribution in an effective Lagrangian approach.  Besides the effective interactions of $B_c\to (c\bar{c})_{J=1} \rho$ as shown in Eq. (\ref{Eq:AmpV1}), we still need the effective Lagrangian for $\rho \pi \pi$, which is \cite{Yan:1992gz,Casalbuoni:1996pg}
\begin{eqnarray}
\mathcal{L}_{\rho\pi\pi}=-ig_{\rho\pi\pi}\rho^+_\mu \big[  \pi^0 {\partial}^\mu\pi^+ -\partial^\mu \pi^0 \pi^+  \big],
\end{eqnarray}
where the coupling constants $g_{\rho\pi\pi}=6.05 $  \cite{Cheng:2004ru}, which is  determined from the $\rho\rightarrow\pi\pi$ partial width.

The $J^P$ quantum numbers of $\zc^{(\prime)}$ are $1^+$, thus $\zc$ couples to $J/\psi \pi$ via $S$-wave, while $\zcp$ couples to $h_c \pi$ via $P$-wave. Considering the chiral symmetry, the effective Lagrangian related to $\zc^{(\prime)}$ are \cite{Liu:2008qx,Huang:2015xud,Chen:2017abq}
\begin{eqnarray}
\mathcal{L}_{Z_c\psi\pi}&=&g_{Z_c\psi\pi}\partial_\mu\psi_\nu (\partial^\mu \pi Z^{\nu}_c-\partial^\nu \pi Z^{\mu}_c) , \nonumber\\
\mathcal{L}_{Z_c^\prime h_c\pi}&=&g_{Z_c^\prime h_c\pi}\varepsilon_{\mu\nu\alpha\beta} \partial^\mu h^\nu_{c} Z^{\prime\alpha}_c \partial^\beta \pi + \mathrm{H.c.},\label{Eq:Lag}
\end{eqnarray}
respectively. The coupling constants in the Lagrangian will be discussed in the next section.

The amplitude of $B^+_c(p)\rightarrow \rho^+(p_1)J/\psi(p_2)[\pi^+(q)]\rightarrow \pi^0(p_3)Z^+_c(p_4)$  corresponding to Fig. \ref{fig:feyn-Zb-1}a is

\begin{eqnarray}
\mathcal{M}_{a}&=&(i)^3\int\frac{d^4 q}{(2\pi)^4}\mathcal{A}^{\mu \nu}(P,Q)
 \big[g_{\rho\pi\pi}(q-p_3)^\xi \big]\nonumber\\ &&\big[-g_{Z_c\psi\pi}p_{2\rho}(q^\rho\epsilon^\sigma_4-q^\sigma\epsilon^\rho_4) \big]
\frac{-g_{\mu\xi}+p_{1\mu}p_{1\xi}/m^2_\rho}{p_1^2-m^2_\rho} \nonumber\\
&&\frac{-g_{\nu\sigma}+p_{2\nu}p_{2\sigma}/m^2_\psi}{p_2^2-m^2_\psi}\frac{1}{q^2-m^2_\pi}
\mathcal{F}(q^2,m^2_\pi),\label{Eq:B1}
\end{eqnarray}
and in a similar way, one can get the amplitude of $B^+_c(p)\rightarrow \rho^+(p_1) h_c(p_2) [\pi^+(q)]\rightarrow \pi^0(p_3)Z^{\prime+}_c(p_4)$ corresponding to orresponding to Fig. \ref{fig:feyn-Zb-1}b, which is,
\begin{eqnarray}
\mathcal{M}_{b}&=&(i)^3\int\frac{d^4 q}{(2\pi)^4}\mathcal{A}^{\mu \nu}(P,Q)
\big[g_{\rho\pi\pi}(q-p_3)^\xi\big] \nonumber\\ &&\big[-g_{Z^\prime_c h_c\pi}\varepsilon^{\rho\sigma\tau\lambda}\epsilon_{4\rho} p_{2\tau}q_\lambda \big]
\frac{-g_{\mu\xi}+p_{1\mu}p_{1\xi}/m^2_\rho}{p_1^2-m^2_\rho} \nonumber\\
&&\frac{-g_{\nu\sigma}+p_{2\nu}p_{2\sigma}/m^2_{h_c}}{p_2^2-m^2_{h_c}}\frac{1}{q^2-m^2_\pi}\mathcal{F}(q^2,m^2_\pi),\label{Eq:B2}
\end{eqnarray}
where $\mathcal{F}(q^2,m_\pi^2) $ is a form factor introduced to depict the structure effects of the exchanged pion meson. In the present work, a monopole form form factor is adopted, which is,
\begin{eqnarray}
\mathcal{F}(q^2,m_\pi^2) =\frac{m_\pi^2 -\Lambda^2}{q^2-\Lambda^2},
\end{eqnarray}
where $\Lambda$ is a parameter, which is of order 1 GeV \cite{Tornqvist:1993vu, Tornqvist:1993ng, Locher:1993cc, Li:1996yn}.

\section{Numerical results and discussion}
\label{sec:results}

Before we estimate the partial width of $B_c^+ \to \zc^{(\prime)+} \pi^0$, we have to discuss the coupling constants  $g_{\zc J/\psi \pi}$ and $g_{\zcp h_c \pi}$. Experimentally, $\zc$ dominantly decay into $D^\ast \bar{D}+c.c$ and $J/\psi \pi$,  and the ratio of these two channels are measured to be $6.2 \pm 1.1 \pm 2.7$. With the center value of $\zc$ width and the assumption that the width of $\zc$ comes from the partial widths of $D^\ast \bar{D}$ and $\pi J/\psi$, we can get the partial width of $\zc \to \pi J/\psi$, and then the estimated coupling constant   $g_{\zc J/\psi \pi}$ is
\begin{eqnarray}
g_{\zc J/\psi \pi}=0.47\ \mathrm{GeV}^{-1}.
\end{eqnarray}
In a similar way, one can estimate the coupling constant $g_{\zcp h_c \pi}$. We suppose $\zcp$ dominantly decay into $D^\ast \bar{D}^\ast$ and $h_c \pi$ and the ratio of the partial widths of these two decay modes were measured to be $12.0 \pm 3.3$ \cite{Ablikim:2013wzq,Ablikim:2013emm}, then with the center values of the width and the ratio, we can get the partial width of $\zcp \to h_c \pi $ and the corresponding coupling constant $g_{\zcp h_c \pi}$, which is
\begin{eqnarray}
g_{\zcp h_c\pi}=0.65~\mathrm{GeV^{-1}}.
\end{eqnarray}

\begin{table}[t]
\begin{center}
\caption{The values of the parameters $F(0)$, $a$ and $b$ in the form factors of $B_c\rightarrow J/\psi$ and $B_c\rightarrow h_c$ \cite{Issadykov:2018myx,Wang:2009mi}. Here, one should notice that these parameters correspond to the definitions in Eqs. (\ref{Eq:tran1}) and (\ref{Eq:tran2})}\label{Tab:PARA1}
  \setlength{\tabcolsep}{3mm}{
\begin{tabular}{cccccc}
  \toprule[1pt]
  $(c\bar{c})$& Parameters &        $A_0$ & $A_+$ & $A_-$ & $V$ \\
  \midrule[1pt]
           & $F(0)$ & 1.65 & 0.55 & -0.87 & 0.78 \\
  $J/\psi$ & $a$ & 1.19 & 1.68 & 1.85 & 1.82 \\
           & $b$ & 0.17 & 0.70 & 0.91 & 0.87 \\
  \midrule[1pt]
             &$F(0)$  & 0.64 & 0.50 & -0.32 & 0.07 \\
  $h_c$    &$a$  & 1.92 & 1.54 & 2.63 & 2.32 \\
           &$b$  & 0.39 & 0.24 & 0.63 & 0.49 \\
  \bottomrule[1pt]
\end{tabular}}
\end{center}
\end{table}

\begin{table}[htb]
\begin{center}
\caption{Values of  parameters $\Lambda_1$, $\Lambda_2$ and $\Lambda_3$. }\label{Table:PARA2}
  \setlength{\tabcolsep}{3mm}{
\begin{tabular}{cccccc}
 \toprule[1pt] 
  $(c\bar{c})$  &   Parameter&   $A_0$ & $A_+$ & $A_-$ & $V$ \\
  \midrule[1pt]
           &$\Lambda_1$ & 10.0 & 9.05 & 7.35 & 7.69 \\
  $J/\psi$ &$\Lambda_2$ & 7.30 & 6.13 & 6.34 & 6.24 \\
           &$\Lambda_3$ & 17.0 & 15.9 & 16.3 & 16.2 \\
  \midrule[1pt]
           &$\Lambda_1$ & 7.02 & 7.88 & 6.05 & 6.43 \\
  $h_c$    &$\Lambda_2$ & 7.02 & 7.88 & 6.05 & 6.41 \\
           &$\Lambda_3$ & 14.0 & 14.0 & 14.0 & 14.0 \\
\bottomrule[1pt]
\end{tabular}}
\end{center}
\end{table}

In the weak interaction vertex, there are four form factors, which are $V(Q^2)$, $A_{0, \pm}(Q^2)$. As indicated in Ref. \cite{Cheng:2003sm}, the form factors are usually estimated in the quark model, thus, the form factors are known only in spacelike region.  Thus, one need to analytically continue the form factors to the timelike region, where the physical decay processes are relevant. In Ref. \cite{Issadykov:2018myx}, the form factors for $B_c\to J/\psi$ are parameterized as the form
\begin{eqnarray}
F(Q^2)=\frac{F(0)}{1-a\zeta+b\zeta^2}.\label{Eq:A1}
\end{eqnarray}
As for $B_c \to h_c$, the form factors are parameterized as \cite{Wang:2009mi}.
\begin{eqnarray}
F(Q^2)=F(0)\exp(a\zeta+b\zeta^2), \label{Eq:A2}
\end{eqnarray}
with $\zeta=Q^2/m^2_{B_c}$ and $F(0)$, $a$ and $b$ are parameters. For completeness of the present work, we collect the values of these parameters in Table \ref{Tab:PARA1}.

\begin{figure}[tb]
  \centering
 \includegraphics[width=8.5cm]{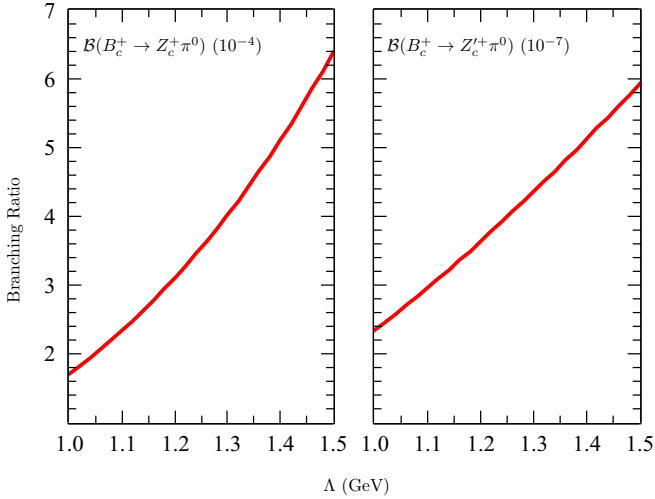}
  \caption{Branching ratios of $B_c\rightarrow Z_c\pi$ (right panel) in unit of $10^{-4}$ and $B_c\rightarrow Z^\prime_c\pi$ (left panel) in unit of $10^{-7}$ depending on parameter $\Lambda$.}\label{Fig:Bc}
\end{figure}

In order to avoid ultraviolet divergence in the loop integrals and evaluate the loop integrals with Feynman parameterization methods, we further parameterize the form factors in the form
\begin{eqnarray}
F(Q^2)= -F(0)\frac{\Lambda^2_1}{Q^2-\Lambda^2_1}\frac{\Lambda^2_2}{Q^2-\Lambda^2_2}\frac{\Lambda^2_3}{Q^2-\Lambda^2_3},\label{Eq:A3}
\end{eqnarray}
where the values of $\Lambda_1$, $\Lambda_2$ and $\Lambda_3$ are obtained by fitting Eqs.~(\ref{Eq:A1}) and (\ref{Eq:A2}) with Eq. (\ref{Eq:A3}) and the fitted parameter values are list in Table~\ref{Table:PARA2}.

In the present work, there is only one model parameter,  $\Lambda$, which is introduced by the form factor in the amplitudes and generally, the magnitude of $\Lambda$ is of order 1 GeV. In Fig. \ref{Fig:Bc}, we present the $\Lambda$ dependence of the branching ratio of $B_c^+ \to \zc^{(\prime)+} \pi^0$. Here, we varies $\Lambda$ from 1 GeV to 1.5 GeV, and in this parameter range, we find the branching ratio of $B_c^+ \to \zc^+ \pi^0$ monotonously increases with the increasing of $\Lambda$. In particular, the branching ratio is predicted to be $(1.71- 6.37 )\times10^{-4}$, which is large enough to be detected in further exclusive $B_c$ decays. This is comparable to the branching ratio of $B_c\to X(3872)\pi$ in a model calculation~\cite{Wang:2007sxa,Wang:2015rcz}.  As for $B_c^+ \to {\zcp}^+ \pi^0$, the branching ratio is estimated to be $(2.34-5.92)\times10^{-7}$, which is about 3 order smaller than the one of $B_c^+ \to \zc^+ \pi^0$.

Our estimation indicates a strong suppression of $\zcp$ production from $B_c$ decay comparing to the case of $\zc$ production.This suppression is partially resulted from the $P$-wave coupling of $\zcp h_c \pi$. Further more,  in the estimation of $B_c \to \zc \pi$, the term related to  $V(Q^2)$ in Eq. (\ref{Eq:AmpV1}) vanishes after performing the loop integral in Eq. (\ref{Eq:B1}). While for $B_c\to \zcp \pi$, only the term related to $V(Q^2)$ survive, which further suppress $\zcp$ production from $B_c^+$ decay.

\section{Summary}
\label{sec:summary}
After the observations of $\zc^{(\prime)}$ by  BES III and Belle Collaborations, the specific properties of these two charmonium-like states have stimulated great interests of theorist and experimentist to reveal their nature. Addition to mass spectrum and decays of the charmonium-like states, searching more production process are also interesting. Besides the electron-positron annihilation, the signal of $\zc$ have been observed in the semi-exclusive $b-$flavored hadron. However, the charmonium-like state $\zc$ were not observed in the $B$ meson decay, such as $B\to K J/\psi \pi\pi$ and $B\to K D^\ast \bar{D}$. Thus, searching for the source of $\zc$ production in the semi-inclusive b-flavor hadron decay will be intriguing.

In present work, we investigate  $\zc^{(\prime)}$ production from $B_c^+$ decay. We find that production process can occur via $(c\bar{c})_{J=1} \rho$ meson loop, which could enhance the rates of $B_c \to \zc^{(\prime)} \pi$ since all the internal mesons could be on-shell. Our estimation shows that the branching ratio of $B_c^+ \to \zc^+ \pi^0$ is of order of $10^{-4}$, which indicates that $\zc$ production from $B_c$ decay could be a important source of $\zc$ production from the $b$-flavored hadron semi-exclusive decay. Such a large branching fraction can also be tested by the exclusive decay of $B_c^+$ by LHCb. As for $\zcp$, we find the branching ratio of $B_c^+ \to Z^{\prime+}_c \pi^0$ is three orders smaller than the one of $B_c^+ \to Z^{+}_c  \pi^0$, which indicate strong suppression of $\zcp$ production from $B_c$ decay.

\section*{Acknowledgement}
\label{sec:acknowledgement}
This work is supported in part by the National Natural Science Foundation of China (NSFC) under Grant Nos. 11775050, 11675091, and 11835015.

\end{document}